\documentclass{sig-alternate}
\usepackage{mathptmx} 

\newcommand{\ignore}[1]{}
\usepackage[pass]{geometry}
\usepackage{fancyhdr}
\usepackage[normalem]{ulem}
\usepackage[hyphens]{url}
\usepackage{hyperref}
\usepackage{color}
\usepackage{soul}
\usepackage{microtype}

\usepackage{setspace}
\usepackage[normalem]{ulem}
\usepackage{graphicx}
\usepackage{subfig}
\usepackage{soul}
\usepackage{color}
\usepackage[linesnumbered,ruled]{algorithm2e}
\usepackage{changepage}
\usepackage{mathtools}
\usepackage{comment}
\DeclarePairedDelimiter\ceil{\lceil}{\rceil}

\usepackage{enumitem}
\usepackage{fancyhdr}
\usepackage{hyperref}
\usepackage{pifont}
\usepackage{array}

\SetCommentSty{mycommfont}
\usepackage{multirow}

\makeatletter
 \def\SOUL@hlpreamble{%
 \setul{}{2.4ex}
 \let\SOUL@stcolor\SOUL@hlcolor
 \SOUL@stpreamble
 }
\makeatother

\newcommand{\arch}{Cache Telepathy}
\soulregister\arch7

\newcommand{\todo}[1]{\hl{\textbf{#1}}}
\newcommand{\mengjia}[1]{\textcolor{red}{\footnotesize{[Mengjia: #1]}}}

\sethlcolor{white}

\fancypagestyle{firstpage}{
  
  \pagenumbering{arabic}
}

\title{\arch: Leveraging Shared Resource Attacks
to Learn DNN Architectures}
\author{Mengjia Yan, Christopher Fletcher, Josep Torrellas\\[3pt]
University of Illinois at Urbana-Champaign\\[3pt]
\{myan8, cwfletch, torrella\}@illinois.edu}

\begin{document}

\maketitle
\thispagestyle{firstpage}
\pagestyle{plain}


\begin{abstract}
\setstretch{1.0}

\hl{Deep Neural Networks (DNNs) are fast becoming ubiquitous for their ability 
to attain good accuracy in various machine learning tasks. A DNN's 
architecture (i.e., its hyper-parameters) broadly determines the DNN's accuracy and 
performance, and is often confidential. Attacking a DNN
in the cloud to obtain its architecture can potentially provide
major commercial value. Further, attaining a DNN's architecture facilitates other, existing DNN attacks.}

\hl{This paper presents }\emph{\hl{\arch}}\hl{: a fast and accurate mechanism to 
steal a DNN's architecture using the cache side channel.
Our attack is based on the insight that DNN inference relies heavily on 
tiled GEMM (Generalized Matrix Multiply), and that DNN architecture 
parameters determine the number of GEMM calls and the dimensions of
the matrices used in the GEMM functions. Such information can be 
leaked through the cache side channel.}

\hl{This paper uses Prime+Probe and Flush+Reload to
attack VGG and ResNet DNNs running
OpenBLAS and Intel MKL libraries. 
Our attack is effective in helping obtain the architectures
by very substantially reducing the search space of target 
DNN architectures. For example, for VGG using OpenBLAS,
it reduces the search space from more than $10^{35}$ 
architectures to just 16.}


%

\end{abstract}

\section{Introduction}
\label{sec:intro}



For the past several years, Deep Neural Networks (DNNs) have increased in popularity 
thanks to  their ability to attain high accuracy and performance 
in a multitude of machine learning tasks --- e.g., image and speech recognition~\cite{SpeechSynth,ResNet}, scene generation~\cite{CNNGan}, and game playing~\cite{AlphaGo}.
An emerging framework that provides end-to-end infrastructure for using DNNs is Machine Learning as a Service (MLaaS)~\cite{AmazonMLaaS,GoogleMLaaS}.
In MLaaS, trusted clients submit DNNs or training data to MLaaS service providers (e.g., an Amazon or Google datacenter).
Service providers host the DNNs, and allow remote \emph{untrusted} users to submit queries to the DNNs for a fee.

\hl{Despite its promise, MLaaS provides new ways to undermine the privacy of  the hosted DNNs.
An adversary may be able to learn details of the 
hosted DNNs beyond the 
official query API. For example, an adversary may try 
to learn the DNN's 
architecture (i.e., its \textit{hyper-parameters}). These are the parameters that give the network its shape, such as the number and types of layers, the number of neurons per layer, and
the connections between layers.}

\hl{The architecture of a DNN broadly determines the 
DNN's accuracy and performance. For this reason,
obtaining it often 
has high commercial value. In addition, it
generally takes a large amount of time and resources to 
try to obtain it by tuning hyper-parameters 
through training. Further, once a DNN's architecture is known, other attacks
are possible, such as the model extraction attack~{\cite{tramer2016stealing}} (which obtains the
weights of the DNN's edges), and the membership inference attack~{\cite{shokri2016membership}},{\cite{membership_general}}
(which determines whether a input was used to train the DNN).}

\hl{Yet, stealing a DNN's architecture is challenging. 
First, DNNs have a multitude of hyper-parameters, which makes 
brute-force guesswork unfeasible.
Further, the DNN design space has been growing with time, which 
is further aggravating the adversary's task.}
%
%


\hl{This paper proves that despite the large search space, attackers
can quickly and accurately recover DNN architectures in the MLaaS 
setting using the cache size channel. Our  
insight is that DNN inference relies heavily on 
tiled GEMM (Generalized Matrix Multiply), and that DNN architecture 
parameters determine the number of GEMM calls and the dimensions of
the matrices used in the GEMM functions. Such information can be 
leaked through the cache side channel.}

\hl{We present an attack that we call }\emph{\hl{\arch}}\hl{.
It is the first cache side channel attack 
on modern DNNs. It targets DNN inference on general-purpose processors, 
which are widely used for inference in existing MLaaS platforms, 
such as Facebook's~{\cite{facebook}} and Amazon's~{\cite{amazon}}.}
\footnote{\hl{Facebook 
currently relies heavily on CPUs for machine learning inference~{\cite{facebook}}. Most instance types provided by Amazon for MLaaS are CPUs~{\cite{amazon}}.}}

\hl{We demonstrate our attack by implementing it 
on a state-of-the-art platform.
We use Prime+Probe and Flush+Reload to attack the VGG and 
ResNet DNNs running OpenBLAS
and Intel MKL libraries. 
Our attack is effective at helping obtain the architectures
by very substantially reducing the search space of target 
DNN architectures. For example, for VGG using OpenBLAS,
it reduces the search space from more than $10^{35}$ 
architectures to just 16.
}

\hl{This paper makes the following contributions:}
\begin{enumerate}[topsep=0pt,itemsep=-1ex,partopsep=1ex,parsep=1ex,leftmargin=2.5ex]
  
  \item \hl{It provides a detailed analysis of the mapping of DNN hyper-parameters to 
  the number of GEMM calls and their arguments.}
  \item \hl{It implements the first cache-based side channel attack to extract DNN 
  architectures on general purpose processors.}
  \item \hl{It evaluates the attack on VGG and ResNet DNNs 
  running OpenBLAS and Intel MKL libraries.} 
\end{enumerate}

\section{Background}
\label{sec:background}


\subsection{Deep Neural Networks}
\label{sec:background_dnn}

\hl{Deep Neural Networks (DNNs) are a class of ML algorithms that use a cascade of multiple layers of nonlinear processing units for feature extraction and transformation~{\cite{lecun2015deep}}.
There are several major types of DNNs in use today, two popular types being fully-connected neural networks (or multi-layer perceptrons) and convolutional neural networks (CNNs).}

\paragraph{\hl{DNN Architecture}}
\hl{The \emph{architecture} of a DNN, also called the \emph{hyper-parameters}, gives the network its shape. 
DNN hyper-parameters considered in this paper are:}
\begin{enumerate}[topsep=2pt,itemsep=-3pt,partopsep=1ex,parsep=1ex,leftmargin=3ex,label=\alph*)]
    \item \hl{Total number of layers.}
    \item \hl{Layer types, such as  fully-connected, convolutional,
      or pooling layer.}
    \item \hl{Connections between layers, including sequential and non-sequential connections such as shortcuts and branches.
    Non-sequential connections exist in recent DNNs, such as ResNet~{\cite{ResNet}}. For example, a shortcut 
    consists of summing up the output of two different layers and using the result as input for a later layer.}
    \item \hl{Hyper-parameters for each layer. For a fully-connected layer, 
    this is the number of neurons in that layer. For a convolutional layer, this is the number of filters, the filter size, and 
    the striding size.}
    \item \hl{The activation function in each layer, e.g., \texttt{relu} and \texttt{sigmoid}.}
\end{enumerate}

\paragraph{\hl{DNN Weights}}
\hl{The DNN \emph{weights}, also called \emph{parameters}, specify operands to multiply-accumulates (MACCs) in the nested function. In a fully-connected layer, each edge out of a neuron is a MACC with a weight; in a 
convolutional layer, each filter is a sliding window that computes dot products over input neurons.}



\paragraph{DNN Usage}
\hl{DNNs usage is two distinct phases: training and inference.
In training, the DNN designer starts with
a network architecture and a training set of labeled inputs, and tries to find the DNN weights to minimize mis-prediction error.
Training is generally performed offline on GPUs and takes a
relatively long time to finish, typically hours or days~{\cite{facebook}},{\cite{de2017understanding}}.
In inference, the trained model is deployed and used to make real-time predictions on new inputs.
For good responsiveness, inference is generally performed on CPUs~{\cite{facebook}},{\cite{amazon}}.

Hyper-parameter tuning~{\cite{hypertune}},{\cite{hypertune2}} is the process of searching for the ideal DNN architecture. 
It requires training multiple DNNs with different architectures on the same data set.
The DNN architecture search space may be large, meaning overall training takes significantly longer than the time to train a single DNN architecture.}



\subsection{Prior Privacy Attacks Need Architecture}
\label{sec:background_ml_privacy}


To gain insight into the importance of DNN architectures,
we discuss prior DNN privacy attacks~\cite{tramer2016stealing,shokri2016membership,membership_general,wang2018stealing}.
\hl{There are three types of such attacks, each with a different goal. All of them require knowing the victim's DNN architecture.
In the following, we refer to the victim's network as the oracle 
network, its architecture as the oracle DNN architecture, and its
training data set as the oracle training data set.

In the \textit{model extraction attack}~{\cite{tramer2016stealing}},
the attacker tries to obtain the weights of the oracle network.
Although this attack is
referred to as a ``black-box'' attack, it assumes that
the attacker knows the oracle DNN architecture at the start. 
The attacker creates a synthetic data set, requests the classification results from the oracle network, and uses such results to train a 
network that uses the oracle architecture.

The \textit{membership inference attack}~{\cite{shokri2016membership}},{\cite{membership_general}} aims to infer the composition of the oracle training data set. This attack also requires knowledge of the oracle DNN architecture.
The attacker creates multiple synthetic data sets and trains multiple networks that use the oracle architecture. Then, he 
runs the inference algorithm on these networks with some inputs in their training sets and
some not in their training sets.  He then compares the results and 
learns the output patterns of data in the training sets. He then
uses this information to analyze the outputs of the oracle network
running the inference algorithm on some inputs, and identifies those
inputs that were in the oracle training data set.

The \textit{hyper-parameter stealing attack}~{\cite{wang2018stealing}} steals the loss function and regularization term
used in ML algorithms, including DNN training and inference.
This attack also relies on knowing the oracle DNN architecture.
During the attack, the attacker leverages the model extraction attack to learn the DNN's weights. He then finds the loss function that
minimizes the training misprediction error.}

\subsection{Cache-based Side Channel Attacks}
\label{sec:background:sidechannel}

Flush+Reload~\cite{yarom2014flushreload} and Prime+Probe~\cite{liu2015practical} are two powerful
cache-base side channel attacks. They can extract
secret keys~\cite{accessAES, everyday} and leak kernel and process information~\cite{meltdown, spectre}.

Flush+Reload requires that the attacker share
secure-sensitive code or data with the victim. This sharing can be achieved by leveraging the page de-duplication technique.
It has been shown that de-duplication is widely deployed in  public clouds to reduce the memory footprint and improve responsiveness~\cite{skarlatos2017pageforge}.
In an attack, the attacker first performs a $\mathsf{clflush}$ 
operation to the shared cache line, to push it out of the cache.
It then waits to allow the victim to execute.
Finally, it re-accesses the same cache line and measures the access latency.
Depending on the latency, it learns whether the victim 
has accessed the shared line.

\hl{Prime+Probe does not require page sharing.
The attacker constructs a collection of addresses, called conflict addresses, which map to the same cache set as the victim's line.
In an attack, the attacker first accesses the 
conflict addresses to cause cache conflicts with the victim's line, and evict it from the cache. After waiting for an interval, it re-accesses the conflict addresses and measures the access latency. The latency is used to infer whether the victim 
has accessed the line.}


\subsection{Threat Model}

This paper develops a cache-timing attack that accurately reveals
a DNN's architecture.
The attack relies on the following standard assumptions.


\paragraph{Co-location}
We assume the attacker process can use techniques from prior work~\cite{getoffcloud,paas} to co-locate onto the same processor chip as the victim process running DNN inference.
\hl{This is feasible, as current MLaaS jobs are deployed on shared clouds. 
Note that recent MLaaS, such as Amazon SageMaker~{\cite{amazon_sagemaker}} and Google ML Engine~{\cite{google_ml_engine}} allow users to upload their own code for training and inference, instead of using pre-defined APIs.
In this case, attackers can disguise themselves as an MLaaS process and the cloud scheduler will be unable to separate attacker processes from victim processes.}


\paragraph{Code Analysis}
We also assume that the attacker can analyze the ML framework code and linear algebra libraries used by the victim.
These are realistic assumptions.
First, open-source ML frameworks are widely used for efficient development of ML applications.
The frameworks supported by Google, Amazon and other companies,
including Tensorflow~\cite{abadi2016tensorflow}, Caffe~\cite{jia2014caffe}, and MXNet~\cite{mxnet} are all public.
Our  analysis is applicable to almost all of these frameworks.
Second, the frameworks' backends are all supported by high-performance and popular linear algebra libraries, such as OpenBLAS~\cite{xianyi2013openblas}, Eigen~\cite{eigen} and MKL~\cite{mkl}.
OpenBLAS and Eigen are open sourced, 
and MKL can be reverse engineered, as we show later.
\hl{While we do not specifically address other algorithms such 
as FFT or Winograd, they are all amenable to 
cache-based attacks similar to those presented here.}

\section{Attack Overview}
\label{challenge}




In this section, we discuss the challenge of reverse-engineering  DNN architectures and our overall attack procedure.

\begin{figure}[ht]
    \centering
    \includegraphics[width=\columnwidth]{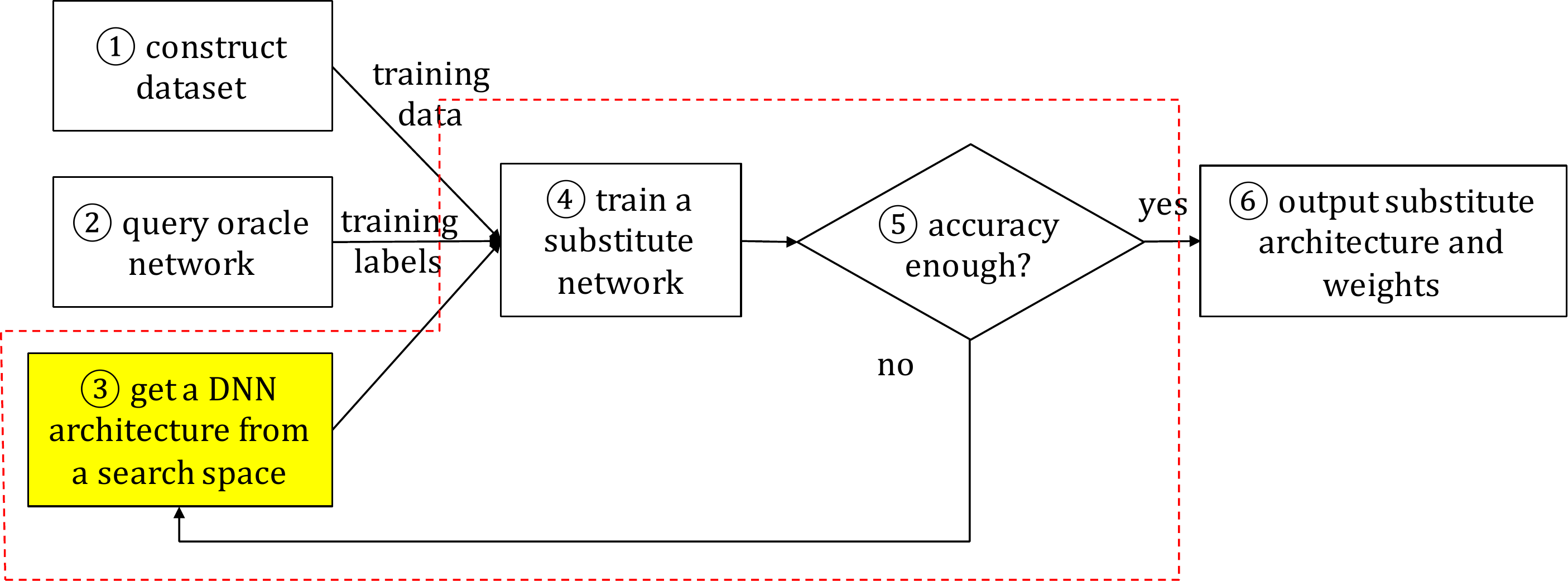}\vspace{-5pt}
    \caption{Searching DNN architectures.}
    \label{fig:extraction}
\end{figure}

\paragraph{Challenge of Reverse-Engineering DNN Architectures}
A DNN's architecture can be extracted in a search process, as shown in Figure~\ref{fig:extraction}.
Specifically, the attacker constructs a synthetic dataset (\ding{172}),
and queries the oracle network for labels or confidence values (\ding{173}), which are used as training data and labels.
Then the attacker chooses a DNN architecture from a search space (\ding{174}) to train a network (\ding{175}).
Steps \ding{174}-\ding{175} repeat until an architecture is found with sufficient prediction accuracy (\ding{176}).

This search process is extremely compute intensive, since it involves many iterations of step \ding{175}.
Considering the depth and complexity of state-of-the-art DNNs, training and validating each network can take hours to days. 
Moreover, without any information about the architecture, the
search space of possible architectures is often intractable. 
Thus, reverse engineering a DNN architecture by
naively searching  DNN architectures is unfeasible.
The goal of \arch\ is to reduce the architecture search space
to a tractable size.

\paragraph{Overall \arch\ Attack Procedure}
We observe that DNN inference relies on GEMM, and that the DNN's hyper-parameters are closely related to the GEMM matrix parameters.
Since high-performance GEMM implementations are tuned for the cache hierarchy through matrix blocking (i.e., tiling), we find that their cache behavior leaks matrix parameters.
In particular, the block size is public (or can be easily deduced), and the attacker can count blocks to learn the matrix sizes.


For our attack, 
we first conduct a detailed analysis of how GEMM is used in ML frameworks, and figure out the mapping between DNN hyper-parameters and matrix parameters (Section~\ref{sec:mapping}).
Our analysis is applicable to most ML frameworks, including TensorFlow~\cite{abadi2016tensorflow}, Caffe~\cite{jia2014caffe}, Theano~\cite{theano}, and MXNet~\cite{mxnet}.

\arch\ includes a cache attack and post processing steps.
First, it uses a cache attack to monitor matrix multiplications and obtain matrix parameters (Section~\ref{sec:attack}
and \ref{sec:mkl}).
Then, the DNN architecture is reverse-engineered based on
the mapping between DNN hyper-parameters and matrix parameters.
Finally, \arch\ prunes the possible values of the remaining undiscovered hyper-parameters and generates 
a pruned search space for the possible DNN architecture (Section~\ref{sec:space}).



\section{Mapping DNNs to Matrix Parameters}
\label{sec:mapping}


DNN hyper-parameters, listed in Section~\ref{sec:background_dnn}, can be mapped to GEMM execution.
We first discuss how the layer type and configurations within each layer map to matrix parameters, assuming that
all layers are sequentially connected (Section~\ref{analysis} and  \ref{resolving}).
\hl{We then generalize the mapping by showing how the connections between layers map to GEMM execution (Section~{\ref{sec:map_shortcut}}).}
Finally, we discuss what information is required to extract the activation functions of Section~\ref{sec:background_dnn} (Section~\ref{activation}).


\subsection{Analysis of DNN Layers}
\label{analysis}


There are two types of neural network layers 
whose computation can be mapped to matrix multiplications, 
namely fully-connected and convolutional layers.

\subsubsection{Fully-connected layer}

In a fully-connected layer, each neuron 
computes a weighted sum of values from all the 
neurons in the previous layer, 
followed by a non-linear transformation.
The $i$th layer computes $\mathsf{out}_i=f_i(\mathsf{in}_i \otimes \theta_i)$ where $\mathsf{in}_i$ is the input vector, 
$\theta_i$ is the weight matrix, 
$\otimes$ denotes a matrix-vector operation,
$f$ is an element-wise non-linear function such as tanh or sigmoid, and $\mathsf{out}_i$ is the resulting output vector.


\begin{table}[h]
    \centering
    \footnotesize
    \begin{tabular}{|c|c|c|}
    \hline
     \textbf{Matrix} & \textbf{n\_rows} & \textbf{n\_columns} \\\hline\hline
     Input: $In_i$ & $B$ & $N_i$\\\hline
     Weight: $\theta_i$ & $N_i$ & $N_{i+1}$\\\hline
     Output: $O_i$ & $B$ & $N_{i+1}$\\\hline
    \end{tabular}
    \caption{Matrix sizes in a fully-connected layer.}\vspace{-5pt}
    \label{tab:fc_map}
\end{table}

The feed-forward computation of a fully-connected DNN is generally performed over a batch of a few  
inputs at a time ($B$).
These multiple input vectors are stacked into an 
input matrix $In_i$.
A matrix multiplication between the input matrix and the weight matrix ($\theta_i$) produces an output matrix, which
is a stack of output vectors. We represent the computation
as $O_i=f_i(In_i \cdot \theta_i)$ where $In_i$ is a matrix with as many rows as $B$ and 
as many columns as $N_i$ (the number of neurons in
the $i$th layer); $O_i$ is a matrix with as many rows as 
$B$ and 
as many columns as $N_{i+1}$ (the number of neurons in
the $i+1$th layer); and $\theta_i$ is a matrix with
$N_i$ rows and $N_{i+1}$ columns. Table~\ref{tab:fc_map}
shows the number of rows and columns in all the matrices.


\subsubsection{Convolutional layer}

\begin{figure}[ht]
    \centering
    \includegraphics[width=.9\columnwidth]{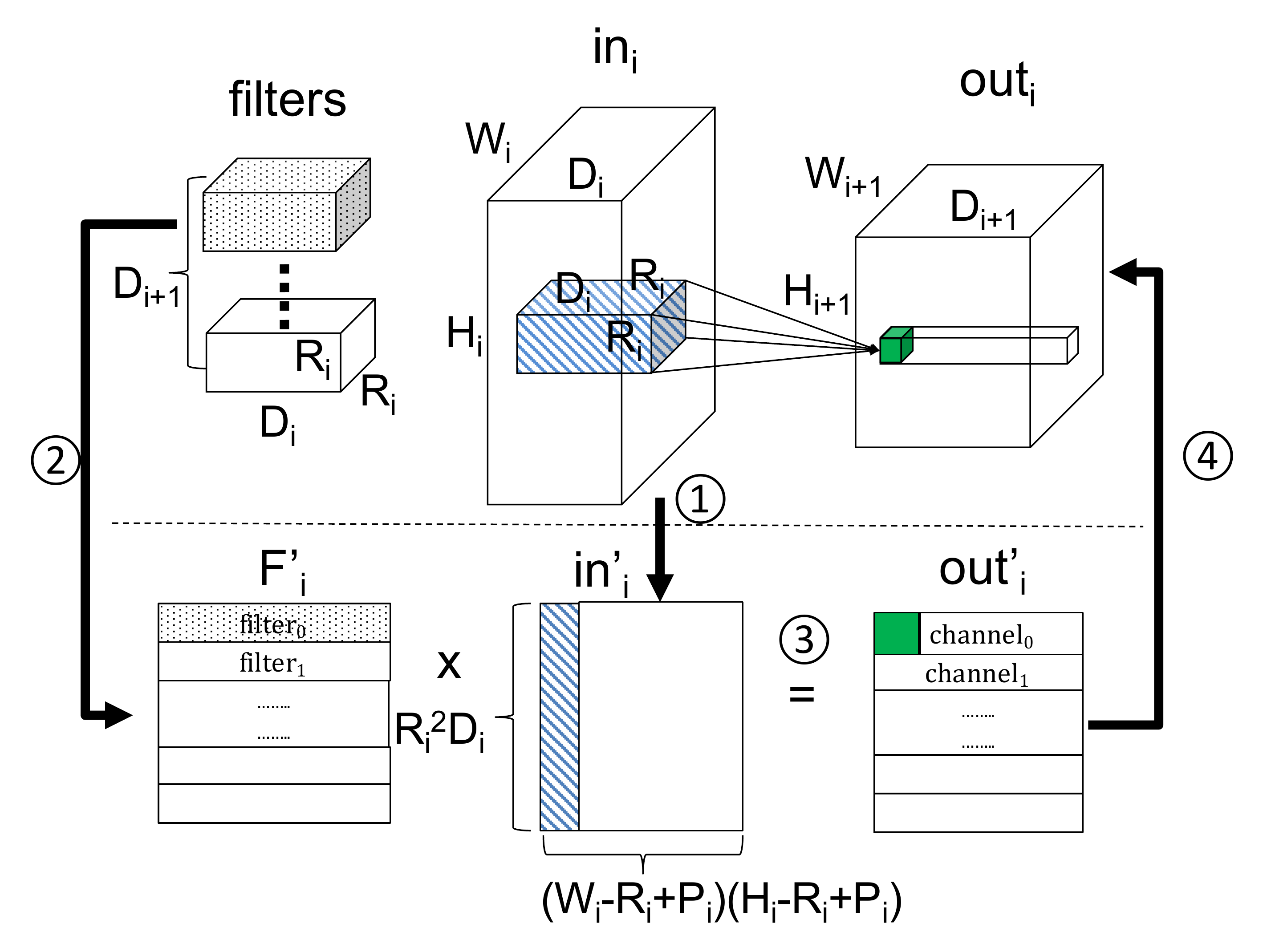}\vspace{-10pt}
    \caption{Mapping a convolutional layer (upper part of the figure) to a matrix multiplication (lower part).}
    \label{fig:conv_layer}
\end{figure}

In a convolutional layer, a neuron is connected to
only a spatial region of neurons in the previous layer.
Consider the upper row of Figure~\ref{fig:conv_layer},
which shows the computation in the $i$th layer. 
The layer generates an output $out_i$ by performing convolution operations on an input image $in_i$
with multiple filters.
The input volume $in_i$ is composed of an image of size $W_i\times H_i$ with $D_i$ channels (center of the
upper row). Each filter is of size $R_i\times R_i\times D_i$ (left part of the upper row).

The figure highlights a neuron in $out_i$ (right part of the upper row). The neuron is a result of a convolution operation --
an elementwise dot product of the filter shaded in dots and the subvolume shaded in dashes.
Both the subvolume and the filter have dimensions $R_i \times R_i \times D_i$.
Applying a filter on the entire input volume ($in_i$) generates one channel of the output ($out_i$).
Thus, the number of filters in $i$th layer ($D_{i+1}$) is the number of channels in the output volume.



The lower row of Figure~\ref{fig:conv_layer} shows a
common implementation that transforms the multiple convolution operations in a layer into a single 
matrix multiply. First, as shown in arrow~\ding{172},
each subvolume in the input volume is stretched out into
a column.
The number of elements in the column is $D_i\times R_i^2$. 
For an input volume with dimensions $W_i\times H_i \times D_i$, 
there are $(W_i-R_i+P_i)(H_i-R_i+P_i)$ such columns in total, where $P_i$
is the amount of zero padding. We call this transformed input matrix $\mathsf{in}_i'$.





Second, as shown in arrow~\ding{173}, individual
filters are similarly stretched out into rows, resulting in matrix $F_i'$. The number of rows in $F_i'$ is the number of filters in the layer.
Then, the convolution becomes a matrix multiply: $\mathsf{out}_i' = F_i' \cdot \mathsf{in}_i'$ (\ding{174} in Figure~\ref{fig:conv_layer}).

Finally, the $\mathsf{out}_i'$ matrix is reshaped back to its proper dimensions
of the $out_i$ volume (arrow~\ding{175}).
Each row of the resulting $\mathsf{out}_i'$ matrix corresponds to one channel in the $out_i$ volume.
The number of columns of the $\mathsf{out}_i'$ 
matrix is $(W_i-R_i+P_i)(H_i-R_i+P_i)$, which is the size of one 
output channel, namely, $W_{i+1}\times H_{i+1}$.

Table~\ref{tab:conv_map}
shows the number of rows and columns in the matrices involved.

\begin{table}[h]
    \footnotesize
    \centering
    \begin{tabular}{|c|c|c|}
    \hline
     \textbf{Matrix} & \textbf{n\_row} & \textbf{n\_column} \\\hline\hline
     $\mathsf{in}_i'$ & $D_i\times R_i^2$ & $(W_i-R_i+P_i)(H_i-R_i+P_i)$\\\hline
     $F_i'$ & $D_{i+1}$ & $D_i\times R_i^2$\\\hline
     $\mathsf{out}_i'$ & $D_{i+1}$ & $(W_i-R_i+P_i)(H_i-R_i+P_i) = W_{i+1}\times H_{i+1}$\\\hline
    \end{tabular}\vspace{-5pt}
    \caption{Matrix sizes in a convolutional layer.}\vspace{-10pt}
    \label{tab:conv_map}
\end{table}
    

The matrix multiplication described above processes a single input. As with fully-connected DNNs, CNN inference typically
consumes a batch of $B$ inputs in a single forward pass. In this case, a convolutional layer performs $B$ matrix multiplications per pass. This is different from fully-connected layers, where the entire batch is computed using only one matrix multiplication.


\subsection{Resolving DNN Hyper-parameters}
\label{resolving}

Based on the previous analysis, we can now map DNN hyper-parameters to matrix operation parameters assuming all layers are sequentially connected.

\subsubsection{Fully-connected networks}

Consider a fully-connected network. Its hyper-parameters are the number of layers, the number of neurons in each layer ($N_i$) and the activation function per layer.
As discussed in Section~\ref{analysis}, the feed-forward computation
performs one matrix multiplication per layer. Hence, 
we extract the number of layers by counting the number of 
matrix multiplications performed. Moreover, 
according to Table~\ref{tab:fc_map}, the number of neurons in layer 
$i$ ($N_i$)
is the number of rows of the layer's weight matrix ($\theta_i$).
Table~\ref{tab:parameters} summarizes the mappings.

\begin{table}[ht]
    \centering
    \footnotesize
    \begin{tabular}{|c|c|c|}
    \hline
     \textbf{Structure}         & \textbf{Hyper-Parameter} & \textbf{Value} \\\hline\hline
     FC network & \# of layers  & \# of matrix muls\\\hline
     FC layer$_i$               &  $N_i$: \# of neurons     &  $n\_row(\theta_i)$ \\\hline \hline
     Conv network               & \# of Conv layers & \# of matrix muls / $B$\\\hline
     Conv layer$_i$             & $D_{i+1}$: \# of filters  &  $n\_row(F_i')$ \\
    \cline{2-3}
                                &  $R_i$:   &  \multirow{2}{*}{$\sqrt{\frac{n\_row(in_i')}{n\_row(out_{i-1}')}}$} \\
                            & filter width/height\footnotemark & \\
                                \cline{2-3}
                                &  $P_i$: padding     &  compare:\\
                                &                     & $n\_col(out_{i-1}'), n\_col(in_i')$ \\\hline
      \hl{Pool$_i$/}               &  \hl{pool/stride}    &  \multirow{2}{*}{\hl{$\approx \sqrt{\frac{n\_col(out_i')}{n\_col(in_{i+1}')}}$}}\\
      \hl{Stride$_{i+1}$} & \hl{width/height} & \\\hline
    \end{tabular}\vspace{-5pt}
    \caption{Mapping between hyper-parameters and matrix parameters.
    FC stands for fully connected.}
    \vspace{-3mm}
    \label{tab:parameters}
\end{table}

\subsubsection{Convolutional networks}

A convolutional network generally consists of four types of layers: convolutional, Relu, pooling, and fully connected. Recall that each 
convolutional layer involves a batch $B$ of matrix multiplications.
We determine $B$ with the following observation: consecutive matrix multiplications will have the same dimensions if they correspond to the same layer in a batch.  We will see that this is the case below.

In a convolutional layer $i$, the hyper-parameters include the 
number of filters ($D_{i+1}$), the filter width or height ($R_i$),
and the padding ($P_i$). We assume that the filter width and height
are the same, which is the common case. Note that
the depth of the input volume ($D_i$) is not considered; it is
an input parameter, obtained from the previous layer. 

From Table~\ref{tab:conv_map}, we see that 
the number of filters ($D_{i+1}$)
is the number of rows of the filter matrix $F_i'$. 
To attain the filter width ($R_i$), we note that
the number of rows of the $in_i'$ matrix is $D_i\times R_i^2$.
Hence, we first need to find $D_i$, which is the
number of output channels in the previous layer.
It can be obtained from the number of rows of the 
$out_{i-1}'$ matrix. Overall, as summarized in
Table~\ref{tab:parameters}, the filter width
is attained by dividing the number of rows of $in_i'$ by 
the number of rows of $out_{i-1}'$ and performing the 
square root. In the case that the $i$th layer is the first 
one, directly connected to the input,
the denominator of this fraction
is the number of channels of the input image, which is public information.

Padding results in a larger input matrix ($in_i'$).
After resolving the filter width ($R_i$), the value of padding 
can be deduced by determining the difference between number 
of columns of the output matrix of layer $i-1$ ($out_{i-1}'$),
which is $W_i\times H_i$, and the number of columns of the
$in_i'$ matrix, which is $(W_i-R_i+P)(H_i-R_i+P)$.


\hl{A pooling layer can be located in-between two convolutional layers.
It down-samples every channel of the input along width and height.
The hyper-parameter in this layer is the pool width and height (assumed to be the same value), 
which can be inferred as follows.
Consider first the (x,y) size of layer $i$, which is 
$W_{i+1}\times H_{i+1}$ (Table~{\ref{tab:conv_map}}), and is given
by the number of columns in matrix $out_i'$.
Then consider the (x,y) size of the input volume in layer 
$i+1$. If the two are the same, there is no pooling layer; otherwise, we expect to see the next (x,y) size reduced by the square of the pool width.
In the latter case, the exact pool dimension can be found using a similar procedure used to determine $R_i$.
Note that non-unit stride results in the same dimension difference,
thus we are unable to distinguish these two.}

Table~\ref{tab:parameters} summarizes the mappings.

\footnotetext{Specifically, we learn the filter spatial dimensions. If the filter isn't square, the search space grows depending on factor combinations (e.g., 2 by 4 looks the same as 1 by 8).  We note that filters in modern DNNs are nearly always square.}

\subsection{Connections Between Layers}
\label{sec:map_shortcut}

\hl{We generalize the above mapping analysis by showing how to map inter-layer connections to GEMM execution.}

\subsubsection{Mapping shortcut/branch connections}

\hl{A branch or shortcut path has two characteristics that can be expressed in GEMM execution.

First, the sink (destination) of a shortcut or the merge point of a branch is mapped to a relatively longer inter-GEMM latency.
Generally, DNNs perform the following operations between two 
consecutive GEMMs: post-processing the current GEMM's output (e.g., batch normalization) and pre-processing the next GEMM's input (e.g., padding and striding).
Therefore, the inter-GEMM latency should be linearly related to the sum of the current layer's output size and the next layer's input size.
However, at the sink, an extra element-wise matrix addition or subtraction needs to be performed, which incurs extra latency between consecutive GEMM calls.

Second, the source of a shortcut or a branch must have the same output dimension as the sink.
This is because a shortcut/branch only connects two layers whose output dimensions match.

We find that these two characteristics are very useful in reducing the architecture search space.}



\subsubsection{Mapping consecutive connections}

\hl{According to the mapping relationships in Table~{\ref{tab:parameters}},
a DNN places several constraints on GEMM parameters for consecutive convolutional layers.
We can leverage these constraints to identify non-sequential connections.

First, since filter width/height must be integer values,
there is a constraint on the number of rows of the input and output matrix sizes between consecutive layers.
Considering the formula used to derive filter width/height in Table~{\ref{tab:parameters}}, if layer $i-1$ and layer $i$ are consecutively connected, the number of rows in $i$th layer's input matrix ($n\_row(in_i)$) must be the product of the number of rows in the $i-1$th layer's output matrix ($n\_row(out_{i-1})$) and square of an integer number.

Second, since pool size and stride size are integer values,
there is another constraint on the number of columns of the input and output matrix sizes between consecutive layers.
According to the formula used to derive pool/stride size,
if layer $i$ and layer $i+1$ are consecutively connected, the number of columns in $i$th layer's output matrix ($n\_col(out_i)$) must be the product of the number of columns in the $i+1$th layer's input matrix ($n\_col(in_{i+1})$) and square of an integer number.

The two constraints above can help us to locate non-sequential connections.
Specifically, if one of these constraints is not satisfied, we are sure the two layers are not consecutively connected.}

\subsection{Activation Functions}
\label{activation}

So far, this section discussed how DNN parameters map to GEMM calls.
Convolutional and fully-connected layers are post-processed by element-wise non-linear functions which do not appear in GEMM parameters.
We can distinguish \texttt{relu} activations from \texttt{sigmoid} and \texttt{tanh} by monitoring a probe address in the \texttt{sigmoid} function using cache attacks.
We remark that nearly all convolutional layers use \texttt{relu} or a close variant~\cite{ReluVariants,Alexnet,simonyan2014vgg,Googlenet,ResNet}.




\section{Attacking Matrix Multiply}
\label{sec:attack}

We now design a side-channel attack to learn matrix multiplication parameters.
Given the mapping from the previous section, this attack will allow us to reconstruct the DNN architecture.

We analyze state-of-the-art BLAS libraries, which 
have extensively optimized blocked matrix multiply for performance. 
Examples of such libraries are OpenBLAS~\cite{xianyi2013openblas}, BLIS~\cite{van2015blis}, Intel MKL~\cite{mkl} and AMD ACML~\cite{acml}.
We show in detail how to extract the desired
information from the GEMM 
implementation in OpenBLAS. 
In Section~\ref{sec:mkl}, we generalize our attack to other BLAS libraries, using Intel MKL as an example.


\subsection{Analyzing GEMM from OpenBLAS}
\label{sec:gemm}

Function \texttt{gemm\_nn} from the OpenBLAS library performs blocked
matrix-matrix multiplication. It computes $C = \alpha A \cdot B + \beta  C$
where $\alpha$ and $\beta$ are scalars, A is an m $\times$ k matrix, 
B is a k $\times$ n matrix, and C is an m $\times$ n matrix.
Our goal is to extract m, n and k.

\begin{figure}[t]
    \centering
    \includegraphics[width=\columnwidth]{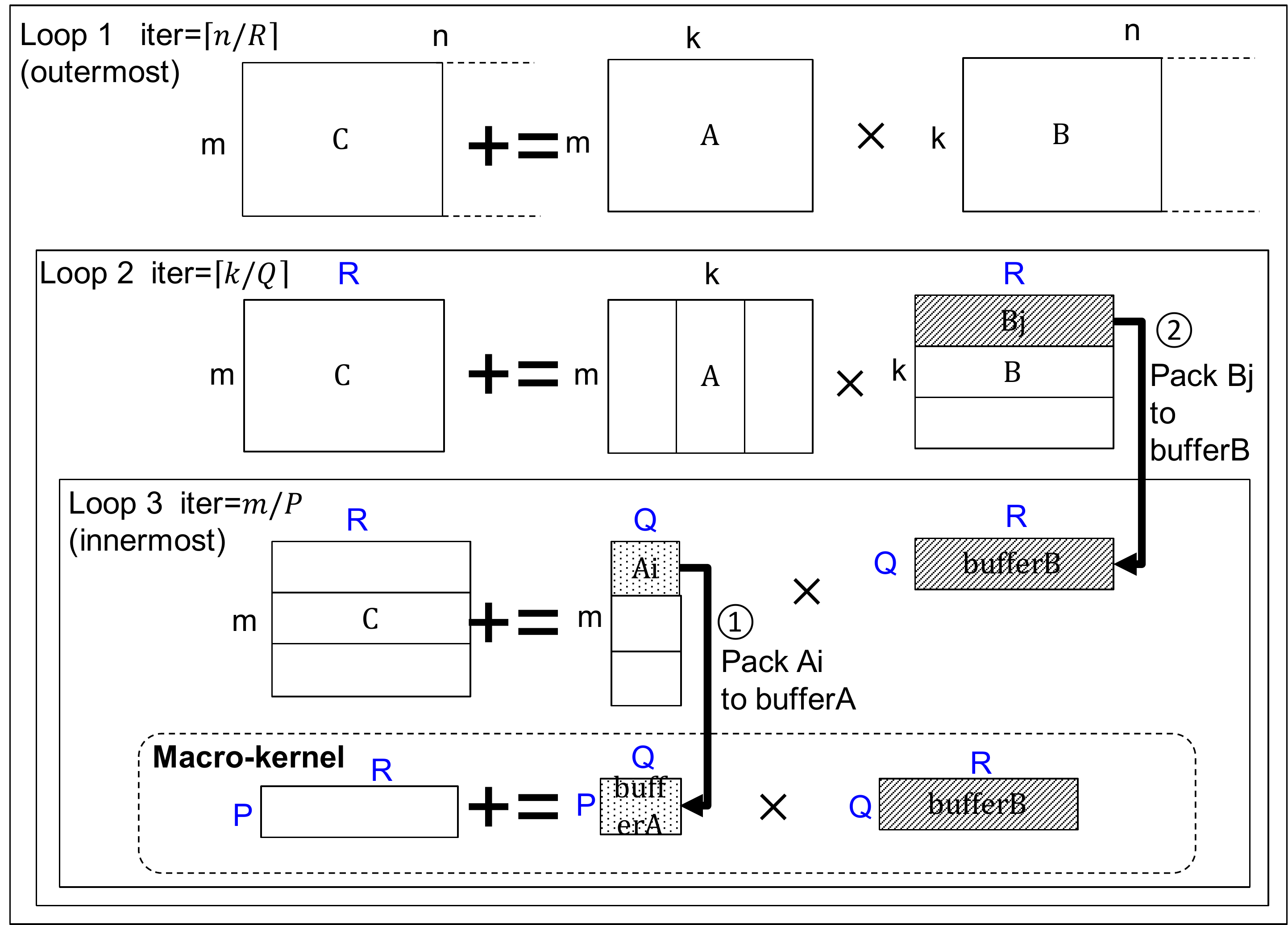}\vspace{-5pt}
    \caption{Blocked GEMM with matrices in column major.}
    \label{fig:gemm}
\end{figure}

Like most modern BLAS libraries, OpenBLAS implements the Goto's algorithm~\cite{goto2008gemm}. Such algorithm has been 
optimized for modern multi-level cache hierarchies and CPUs.
Figure~\ref{fig:gemm} depicts the way the Goto's algorithm structures blocked matrix multiplication for a three-level cache. 

The {\em macro-kernel} at the bottom performs the basic operation,
multiplying a P $\times$ Q block from matrix A with 
a Q $\times$ R block from matrix B. This kernel is generally written in assembly code, and manually optimized by taking CPU pipeline structure and register availability into consideration. The block sizes are picked so that the P $\times$ Q block of A fits in the L2 cache, and the 
Q $\times$ R block of B fits in the L3 cache.

As shown in Figure~\ref{fig:gemm}, there is a three-level loop
nest around the macro-kernel. The innermost one is Loop 3, the
intermediate one is Loop 2, and the outermost one is
Loop 1.
We call the iteration counts in these loops
$iter_3$, $iter_2$, and $iter_1$, respectively, and are given by: 
\begin{equation}
iter_3=\ceil{m/P};\quad iter_2=\ceil{k/Q};\quad iter_1=\ceil{n/R}
\label{eq:gemm}
\end{equation}
Algorithm~\ref{algo:openblas_gemm} shows the corresponding pseudo-code with the three nested loops.
Note that, Loop 3 is further split into two parts, to obtain better cache locality.
The first part performs only the first iteration, and the second part performs the rest.

\begin{algorithm}[t]
\footnotesize
\SetKwInOut{Input}{Input}
\SetKwInOut{Output}{Output}
\Input{Matrix $A$, $B$, $C$; Scalar $\alpha$, $\beta$; Block size $P, Q, R$; $UNROLL$}
\Output{$C := \alpha A \cdot B + \beta C$}
\For(\tcp*[h]{Loop 1}){$j=0, n, R$}{
    \For(\tcp*[h]{Loop2}){$l=0, k, Q$}{
        $itcopy(A[0,l], buf\_A, P, Q)$\\
        \tcp{Loop 3, 1st iteration}
        \For{$jj = j, j+R, 3UNROLL$}{
            $oncopy(B[l, jj], buf\_B+(jj-j)\times Q, Q, 3UNROLL)$\\
            $kernel(buf\_A, buf\_B+(jj-j)\times Q, C[l,j], P, Q, 3UNROLL)$
        }
        \For(\tcp*[h]{Loop 3, rest iterations}){$i=P, m, P$}{
            $itcopy(A[i,l], buf\_A, P, Q)$\\
            $kernel(buf\_A, buf\_B, C[l,j], P, Q, R)$
        }
    }
}
\caption{\texttt{gemm\_nn} in OpenBLAS}
\label{algo:openblas_gemm}
\end{algorithm}

In all the iterations of Loop 3, the data in the P $\times$ Q block from matrix A is packed into a buffer (function \texttt{itcopy}),
before calling the macro-kernel (function \texttt{kernel}).
This is shown in Figure~\ref{fig:gemm} as arrow~\ding{172} and corresponds to Line 3 and 9 in Algorithm~\ref{algo:openblas_gemm}. 
Additionally, in the first part of Loop 3, 
the data in the Q $\times$ R block from matrix B is also packed into a buffer (function \texttt{oncopy}).
This is shown in Figure~\ref{fig:gemm} as arrow~\ding{173} and corresponds to Line 5 in Algorithm~\ref{algo:openblas_gemm}.
The Q $\times$ R block from matrix B is copied in units of Q $\times$ {\em 3UNROLL} sub-blocks.
This breaks down the first part of Loop 3 into a loop with
an iteration count of $iter_4$, given by:
\begin{equation}
\begin{split}
   & iter_4 = \ceil{R/3UNROLL} \\
or\quad & iter_4 = \ceil{(n \bmod R)/3UNROLL}
\end{split}
\label{eq:iter4}
\end{equation}
\noindent where the second expression corresponds to the last
iteration of Loop 1.

\subsection{Locating Probing Addresses}
\label{probing}

Our goal is to find the size of the matrices of Figure~\ref{fig:gemm},
namely, {\em m}, {\em k}, and {\em n}. To do so, we need 
to first obtain the number of 
iterations of Loops 1, 2, and 3, and then use Equation~\ref{eq:gemm}.
Note that we know the values of the block sizes {\em P}, {\em Q}, and 
{\em R} (as well as {\em UNROLL}) --- these are constants
available in the open-source code of OpenBLAS.

A straight-forward approach to obtain the number of iterations of 
Loops 1, 2, and 3 is to monitor the addresses that hold the instructions 
of the loop entries. These are 
Lines 1, 2, and 8 in Algorithm~\ref{algo:openblas_gemm}, respectively. 
We could count the number of times these instructions are executed.
However, this fails because the loop body is very tight. 
Specifically, the instructions for Lines 1 and 2 fall into the same 
cache line, and Line 8 falls into a cache line that is very close to them.
To disambiguate all the loops, we need to monitor better addresses.

\begin{figure}[ht]
    \centering
    \includegraphics[width=\columnwidth]{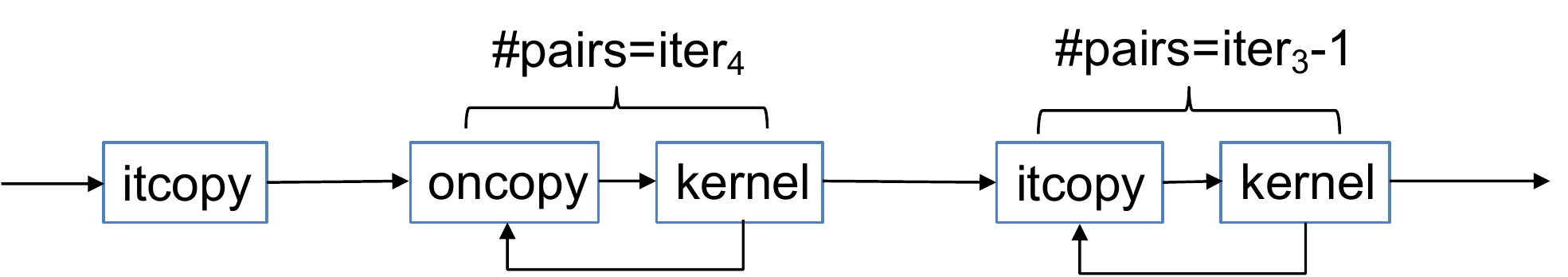}\vspace{-5pt}
    \caption{DCG of \texttt{gemm\_nn}, with the number of invocations per iteration of Loop 2.}
    \label{fig:cfg}
\end{figure}

In this paper, we propose to use, as probing addresses, addresses in
the \texttt{itcopy}, \texttt{oncopy} and \texttt{kernel} functions of
Algorithm~\ref{algo:openblas_gemm}. To understand why, 
consider the dynamic invocations to these functions.
Figure~\ref{fig:cfg} shows 
the Dynamic Call Graph (DCG) of \texttt{gemm\_nn} in Algorithm~\ref{algo:openblas_gemm}. 
Each iteration of 
Loop 2 contains one invocation of function \texttt{itcopy}, 
followed by $iter_4$ invocations of the pair \texttt{oncopy} and \texttt{kernel}, and then $(iter_3-1)$ invocations of the pair \texttt{itcopy} and \texttt{kernel}. 
The whole sequence in Figure~\ref{fig:cfg} is executed 
$iter_1\times iter_2$ times in one invocation of \texttt{gemm\_nn}.

\begin{figure*}[t]
    \centering
    \includegraphics[width=\textwidth]{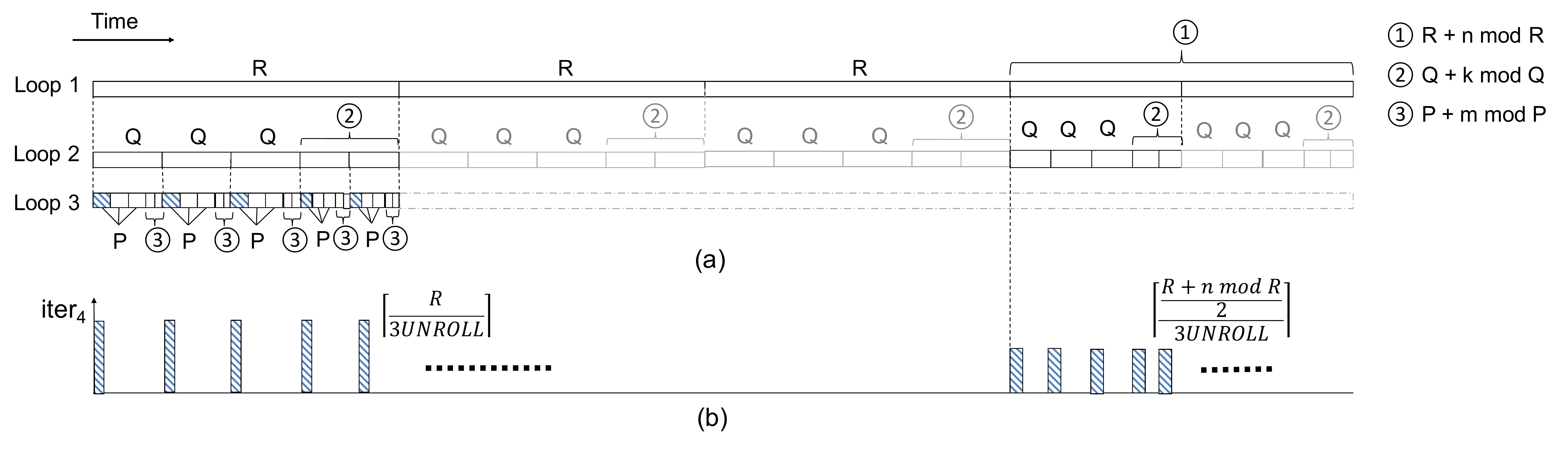}\vspace{-10pt}
    \caption{Visualization of execution time of \texttt{gemm\_nn} where Loop 1, Loop 2, and Loop3 have 5 iterations each (a);
     value of $iter_4$ for each first-iteration of Loop 3 (b).}\vspace{-.5em}
    \label{break}
\end{figure*}

We will see in Section~\ref{sec:attack:post} that these 
invocation counts are enough to allow us to find the size of the matrices of Figure~\ref{fig:gemm}. However, using the first instruction 
in \texttt{itcopy}, \texttt{oncopy}, and \texttt{kernel} as probe
addresses is not optimal. Indeed, we have found that these instructions
are sometimes accessed by prefetchers in the shadow of a branch 
misprediction. This introduces noise in the measurements.

Alternatively, we use one instruction 
inside each function for the three probe addresses. Further, 
since the main bodies of these functions
are loops, we use instructions that are part of the loop (to distinguish it from the GEMM loops, we call it in-function loop).
This improves our monitoring capability, as such instructions are
accessed multiple times per function invocation.


Overall, instructions in the bodies of these three functions
satisfy the conditions for good probing addresses. First,
we will see that the number of accesses to these instructions
can be used to deduce the iteration count of each level of the loop.
Second, these addresses are distant from each other and, hence,
automatic prefetching of instructions does not introduce noise.
Finally, since each function operates on a block of data, 
the intervals between consecutive invocations of these functions 
are long enough.

Note that, even though this DCG is specific for \texttt{gemm\_nn} in
OpenBLAS, it captures two common features for general blocked matrix multiplication. First, all implementations use a kernel function as 
a unit to carry out block-size computation. Second, there are always packing operations before kernel execution at two different levels of the loop. 
A similar DCG can be extracted for slightly different implementations, 
such as Intel's MKL (Section~\ref{sec:mkl}).

\subsection{Procedure to Extract Matrix Dimensions}
\label{sec:attack:post}

To understand the procedure we use to extract matrix dimensions,
we note the way OpenBLAS groups blocks into iterations. Specifically,
rather than assigning a small block to the last iteration, 
it assigns two equal-sized small blocks to the last two iterations. For example,
as it blocks the {\em n} columns of matrix C in Figure~\ref{fig:gemm} into blocks of size
{\em R}, it assigns {\em R} columns to each Loop 1 iteration, except for the last two
iterations. Each of the latter receives {\em (R + n mod R)/2}.

Figure~\ref{break}(a) shows the visualization of the execution time of
\texttt{gemm\_nn} where Loop 1, Loop 2, and Loop 3 have 5 iterations each. 
It shows the size of the block each iteration operates on. In Loop 1, the
first three iterations use {\em R}-sized blocks; each of the last two
use a block of size {\em (R + n mod R)/2}. In Loop 2, the corresponding
block sizes are {\em Q} and {\em (Q + k mod Q)/2}. In Loop 3, they are
{\em P} and {\em (P + m mod P)/2}.

Recall that the first iteration of every Loop 3 invocation is special. While
it processes a {\em P}-sized block, its execution time is different because
it performs a different operation. Specifically, as shown in Figure~\ref{fig:cfg},
it invokes the \texttt{oncopy}-\texttt{kernel} pair $iter_4$ times. 
Figure~\ref{break}(b) shows the value of $iter_4$ for each of the first iterations of Loop 3.
As indicated in Equation~\ref{eq:iter4}, during the execution of ``normal'' Loop 1
iterations, it is $\ceil{R/3UNROLL}$. However, in the last two iterations of
Loop 1, $iter_4$ is $\ceil{((R + n mod R)/2)/3UNROLL}$.

Based on these insights, our procedure to extract {\em m}, {\em k}, and {\em n}
has four steps.

\textbf{Step 1: Identify the DCG of Loop 2 iterations, and 
extract $iter_1\times iter_2$.} 
By probing one instruction in each of \texttt{itcopy}, \texttt{oncopy}, and \texttt{kernel}, 
we repeatedly obtain the DCG pattern of Loop 2 iterations (Figure~\ref{fig:cfg}). 
By counting the number of such patterns, we obtain $iter_1\times iter_2$.

\textbf{Step 2: Extract $iter_3$ and determine the value of $m$.}
In the DCG pattern of a Loop 2 iteration, we count the number of
invocations to the \texttt{itcopy}-\texttt{kernel} pair (Figure~\ref{fig:cfg}).
This count plus 1 gives $iter_3$. Of all these  $iter_3$ iterations, all but the last
two execute a block of size {\em P}; the last two execute a block of size
{\em (P + m mod P)/2} each (Figure~\ref{break}(a)). To estimate the size of this smaller
block, we assume that the execution time of an iteration is proportional
to the block size it processes --- except for the first iteration which, as we
indicated, is different. Hence, we time the execution of a ``normal'' iteration of loop
L3 and the execution of the last iteration of Loop 3. Let's call the times $t_{normal}$
and $t_{small}$. The value of $m$ is:
\vspace{-5pt}\[ m = (iter_3 - 2)\times P + 2\times \frac{t_{small}}{t_{normal}}\times P \vspace{-2pt}\]

\textbf{Step 3: Extract $iter_4$, $iter_2$ and determine the value of $k$.}
In the DCG pattern of a Loop 2 iteration (Figure~\ref{fig:cfg}), we count the number of
\texttt{oncopy}-\texttt{kernel} pairs, and obtain $iter_4$.
As shown in Figure~\ref{break}(b), the value of $iter_4$ is $\ceil{R/3UNROLL}$
in all iterations of Loop 2 except those that are part of the last two iterations of
Loop 1. For the latter, $iter_4$ is $\ceil{((R + n mod R)/2)/3UNROLL}$, which is a lower value.
Consequently, by counting the number of DCG patterns that have a low value of $iter_4$,
and dividing it by 2, we attain $iter_2$. We then follow the procedure of Step 2 to calculate
{\em k}. Specifically, all Loop 2 iterations but the last two execute a block of size 
{\em Q}; the last two execute a block of size
{\em (Q + k mod Q)/2} each (Figure~\ref{break}(a)). Hence, we time the execution of a 
``normal'' iteration and the last one, and compute $k$ as per Step 2.

\textbf{Step 4: Extract $iter_1$ and determine the value of $n$.}
If we count the total number of DCG patterns in the execution and divide that by $iter_2$,
we obtain $iter_1$. We know that all Loop 1 iterations but the last two execute a block of size 
{\em R}; the last two execute a block of size
{\em (R + n mod R)/2} each. To compute the latter, we note that, in the last two iterations
of Loop 1, $iter_4$ is $\ceil{((R + n mod R)/2)/3UNROLL}$. Since both $iter_4$ and $3UNROLL$
are known, we can compute (R + n mod R)/2. Hence, the value of $n$ is:
\[ n = (iter_1 - 2)\times R + 2\times iter_4\times 3UNROLL\]

However, our attack cannot handle the case when
the matrix dimension size is less than twice the block size --- for 
example, when {\em n} is less than $2\times R$. In this case,
there is no iteration that works on a full block. Our procedure cannot
compute the exact value of {\em n}, and can only provide a range of values.

\section{Generalization of the Attack}
\label{sec:mkl}

Our attack can be generalized to other BLAS libraries, since all of them use blocked matrix-multiplication, and most of them implement Goto's algorithm~\cite{goto2008gemm}.
Even though they may differ in the scheduling of the three-level nested loop, block sizes and implementation of the macro-kernel,
our attack is still effective.
In this section, we show that the same attack strategy can be applied to other BLAS libraries, using Intel MKL as an example.
We choose MKL for two reasons. 
First, it is another heavily used library (similar to  OpenBLAS). 
Second, it is closed source, which makes the attack more challenging.

Since MKL's source code is not disclosed, we need to complete two tasks 
before extracting matrix parameters.
First, we need to construct the DCG of GEMM using packing and kernel functions.
This requires us to find the proper functions and their invocation patterns.
Second, we need to obtain the block size of each dimension.
In both tasks, we leverage side channel attacks to assist program analysis.

\paragraph{Constructing the DCG}
First, we use GDB to manually analyze the binary code.
We trace down the functions that are called during matrix multiplication, which we find are the packing functions \texttt{copybn}, \texttt{copyan}.
We find that the kernel function is called \texttt{ker0}.

Next, we determine function invocation patterns and correlate those patterns with loop executions.
This can be achieved in multiple ways.
One possible approach is manual analysis, that is, using GDB to trace the dynamic execution path of GEMM 
and observe the functions invoked for each loop.
Another approach is leveraging cache-based side channel attacks
to probe the packing and kernel functions, and obtain the invocation patterns.

\begin{figure}[ht]
    \centering
    \includegraphics[width=\columnwidth]{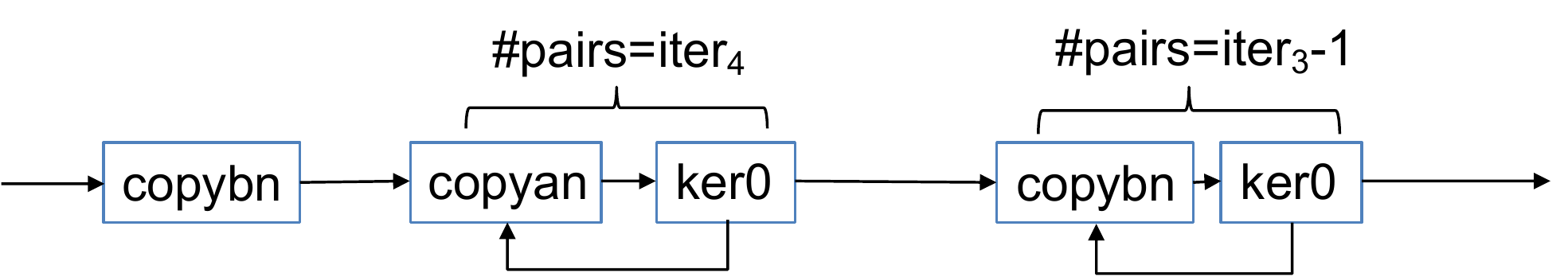}\vspace{-5pt}
    \caption{DCG of blocked GEMM in Intel MKL, with the number of invocations per iteration of Loop 2.}
    \label{fig:cfg_mkl}
\end{figure}

We derive the same DCG using the two approaches, shown in Figure~\ref{fig:cfg_mkl}. The pattern is the same as the DCG of OpenBLAS in Figure~\ref{fig:cfg}.
Thus, the attack strategy in Section~\ref{sec:attack} should also work towards MKL.

\paragraph{Extracting block sizes}
According to Formulas~\ref{eq:gemm} and ~\ref{eq:iter4}, there is a discrete linear relationship between the matrix size and 
the iteration count. We leverage the side-channel attacks in Section~\ref{sec:attack} to count the number of iterations and resolve block sizes.
Specifically, we gradually increase the input dimension size until the number of iterations increments.
For each dimension, the stride on the input dimension that triggers the change of iteration count is the block size.
By applying this approach, we can successfully derive block sizes, which match the sizes we obtain via manual analysis of the MKL binary code.

\paragraph{\hl{Special cases}}
\hl{According to our analysis, MKL follows a different DCG when dealing with small matrices.
Instead of doing 3-level nested loops as in Figure~{\ref{fig:cfg_mkl}}, it uses a single-level loop, tiling on the dimension that has the largest value among $m$, $n$, $k$. The computation is done in place, without triggering packing functions.

For these special cases, we slightly adjust the attack strategy in Figure~{\ref{break}}.
We use side channels to monitor the number of iterations on that single-level loop and the time spent for each iteration.
We then use the number of iterations to deduce the size of the largest dimension. Finally, we use the timing information for each iteration to deduce the product of the other two dimensions.
}


\section{Experimental Setup}
\label{sec:setup}

\paragraph{Attack Platform}
We evaluate our attacks on a Dell workstation Precision T1700, which has a 4-core Intel Xeon E3 processor
and an 8GB DDR3-1600 memory.
The processor has two levels of private caches and a shared last level cache.
The first level caches are a 32KB instruction cache and a 32KB data cache.
The second level cache is 256KB. 
The shared last level cache is 8MB.
We test our attacks on a same-OS scenario using Ubuntu 4.2.0-27.
Our attacks should be applicable to other platforms,
as the effectiveness of Flush+Reload and Prime+Probe has been proved in multiple hardware
platforms~\cite{liu2015practical,yarom2014flushreload}.

\paragraph{\hl{Victim DNNs}}
\hl{We use
a VGG~{\cite{simonyan2014vgg}} instance and a ResNet~{\cite{ResNet}} instance as victim DNNs.
VGG is representative of early DNNs (e.g., AlexNet~{\cite{Alexnet}} and LeNet~{\cite{lenet}}).
ResNet is representative of state-of-the-art DNNs.
Both are standard and widely-used CNNs with a large number of layers and hyper-parameters. 
ResNet additionally features shortcut/branch connections (Section~{\ref{sec:map_shortcut}}).

There are several versions of VGG, with 11 to 19 layers.
All VGGs contain 5 blocks and, within each block, a single 
type of layer is replicated.
We  show our results on VGG-16.

There are several versions of ResNet, whose depth varies from 18 to 152 layers.
All of them consist of the same 4 types of modules, which are replicated a different number of times. 
Each module contains 3 or 4 layers, which are all different.
We  show our detection results on ResNet-50.

The victim programs are implemented using the Keras~{\cite{keras}} framework, with Theano~{\cite{theano}} as the backend.}


\section{Evaluation}
\label{sec:eval}

\hl{We first evaluate our attacks on the GEMM function.
We then show the effectiveness of our attack on neural network inference.
We provide a detailed analysis of extracting the DNN architecture for a VGG-16 and a ResNet-50 instance, and quantify the reduction in the architecture search space.}

\subsection{Attacking GEMM Using Prime+Probe}
\label{sec:eval_primeprobe}

\hl{We show the results of attacking GEMM using Prime+Probe,
which does not require page sharing or the use of the \texttt{clflush} instruction.
Our attack targets the LLC. Hence, victim and attacker can run on different cores, further improving the attack's flexibility.  We locate the LLC sets of two instructions from 
\texttt{itcopy} and \texttt{oncopy}, and construct two sets of eviction addresses using the algorithm proposed by Liu et al.~{\cite{liu2015practical}}.}

\begin{figure}[ht]
    \centering
    \includegraphics[width=\columnwidth]{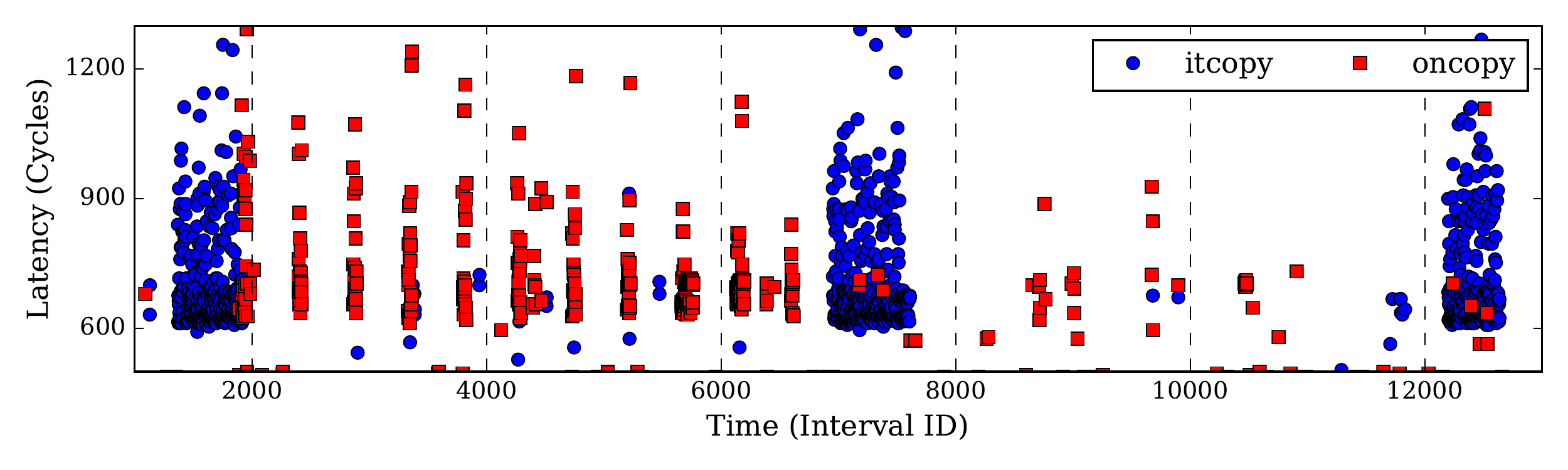}\vspace{-10pt}
    \caption{\hl{Prime+Probe trace of the GEMM execution with monitoring interval length equal to 2000 cycles.}}
    \label{fig:primeprobe}
\end{figure}

\hl{Figure~{\ref{fig:primeprobe}} shows one raw trace of the execution of one iteration of Loop 2 in Algorithm~{\ref{algo:openblas_gemm}}.
We only show the latency above $500$ cycles, which indicates victim activities.
Note that, since we select the probe addresses to be 
within loop bodies (Section~{\ref{probing}}), a cluster of misses marks the time period when the victim is executing the probed function.
In this trace, the victim calls \texttt{itcopy} around interval 2000, then calls \texttt{oncopy} $11$ times between intervals 2000 and 7000.
It then calls \texttt{itcopy} another two times in intervals
7000 and 13000.
The trace matches the DCG shown in Figure~{\ref{fig:cfg}}, and
we can derive that $iter_4=11$ and $iter_3=3$.}

\hl{The noise, such as between intervals 8000 and 12000, can be trivially distinguished from the actual victim accesses.
The trace can be further cleaned up by leveraging de-noising techniques~{\cite{liu2015practical}},{\cite{yarom2014flushreload}}.
As Prime+Probe can effectively extract matrix parameters and achieve the same high accuracy as Flush+Reload, 
either can be used to extract DNN hyper-parameters.
Thus, in the rest  of this section, we use Flush+Reload to illustrate our attack.}

\subsection{Extracting Parameters from DNNs}
\label{sec:eval_nn}

\begin{figure}[t]
    \hspace{-5ex}\includegraphics[width=1.2\columnwidth]{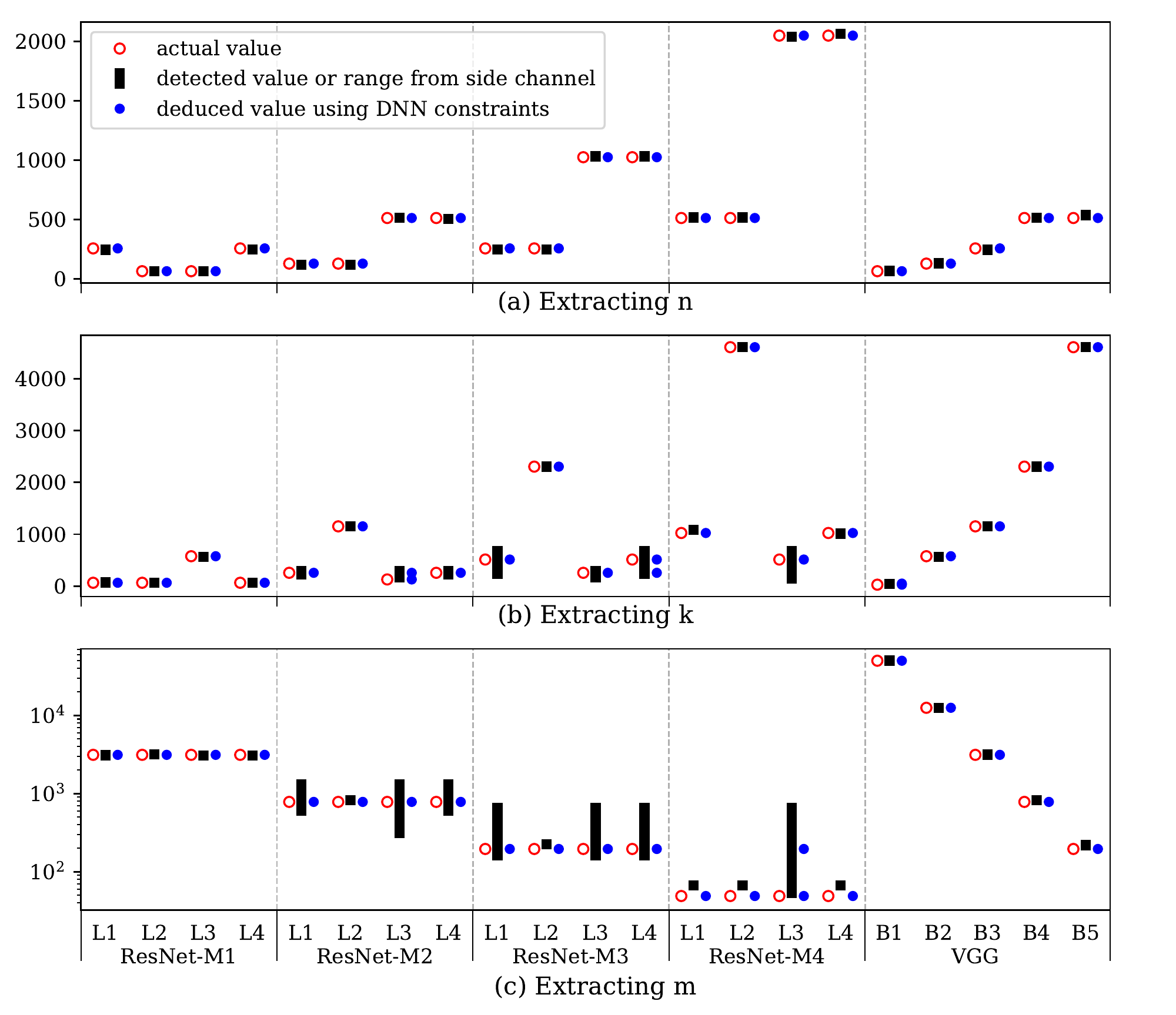}\vspace{-10pt}
    \caption{\hl{Extracted values of the $n$, $k$, and $m$
      matrix parameters for VGG-16 and ResNet-50.}
      }
    \label{fig:dnn_gemm}
\end{figure}

\hl{We show the effectiveness of our attack by extracting the hyper-parameters of our VGG~{\cite{simonyan2014vgg}} and  ResNet~{\cite{ResNet}} instances.
Figure~{\ref{fig:dnn_gemm}} shows the extracted values of the
$n$, $k$, and $m$
matrix parameters for the layers in the 4 distinct modules in ResNet-50, and for the first layer in each of the 5 blocks in VGG-16. We do not show the other layers because they are duplicates of the
layers shown. Figures~{\ref{fig:dnn_gemm}}(a), (b), and (c) correspond
to the values of $n$, $k$, and $m$, respectively.}

\hl{In Figure~{\ref{fig:dnn_gemm}}, for a given parameter (e.g., $m$)
and a given layer (e.g., L1 in ResNet-M2), the hollowed 
circles indicate the \emph{actual}
value of the parameter.
The squares or rectangles indicate
the values of the parameters
\emph{detected} with the side channel attack. When the 
side channel attack can only narrow down the possible values to 
a range, the figure shows a rectangle.
Finally, the solid circles indicate the values of the 
parameters that we \emph{deduce}, using the detected values 
and some DNN constraints.
For example, for parameter $m$ in layer L1 of ResNet-M2,
the actual value is 784, the detected value range is 
524-1536, and the deduced value is 784.}

\hl{We will discuss how we obtain the solid circles later.
Here, we focus on comparing the actual and detected values 
(hollowed circles and squares/rectangles).
Figure~{\ref{fig:dnn_gemm}}(a) shows that our attack is always able to determine the $n$ dimension size with ignorable error, thanks to the small subblock size used in the 1st iteration of Loop 3 (Algorithm~{\ref{algo:openblas_gemm}}).}

\hl{Figures~{\ref{fig:dnn_gemm}}(b) and (c) show that the
attack is  able to accurately determine the $m$ and $k$ values 
for most layers in ResNet-M1, ResNet-M4 and all blocks in VGG.
However, it can only derive ranges of values for most of the 
ResNet-M2 and 
ResNet-M3 layers. This is because the $m$ and $k$ values 
in these layers are often smaller than twice the 
corresponding block sizes (Section~{\ref{sec:attack:post}}).}

\hl{In summary, our side channel attack can either detect the
matrix parameters with negligible error, or can provide a range 
where the actual value falls in.
We will later show that the imprecision from the negligible error and the 
ranges can be eliminated after applying DNN constraints.} 

\subsection{Size of Architecture Search Space}
\label{sec:space}

\hl{The goal of \arch is to very substantially
reduce the search space of possible architectures that
can match the oracle architecture. In this section,
we compare the number of architectures in
the search space without \arch\ (which we 
call \emph{Original} space), and with \arch.
In both cases, we initially reduce the search space by only
considering reasonable hyper-parameters for the layers.
Specifically, for convolutional layers, 
the number of filters can be a multiple of $64$ ($64\times i$, where $1\leq i \leq 32$),
and the filter size can be an integer value between $1$ and $11$.
For fully-connected layers, the number of 
neurons can be $2^i$, where $8 \leq i \leq 13$.}

\subsubsection{Size of the original search space}

\hl{To be conservative, we assume that the attacker knows the number of layers and type of each layer in the oracle DNN.
There exist $352$ different configurations for each convolutional layer without considering pooling or striding,
and $6$ configurations for each fully-connected layer.
Moreover, considering the existence of branches,
given $L$ layers, there are $L\times 2^{L-1}$ 
possible ways to connect them.}

\hl{If we do not count potential shortcuts or branches, a network with 13 convolutional layers and 3 fully-connected layers such as VGG-16 has a search space size of about $4\times10^{35}$. 
If we consider non-sequential connections,  for a network module with only 4 convolutional 
layers such as Module 1 in ResNet-50, there are around $5\times10^{11}$ possible architectures.
Considering that ResNet-50 has 4 different modules, the total search space will be over  $5.8\times10^{46}$.
The search space would be even larger in an actual attack scenario,
since the attacker would not have any information on the total number of layers and layer types.}

\hl{Overall, we can see that the search space is intractable.
Since training each candidate DNN architecture takes hours to days,
relying on the original search space does not lead to
obtaining the oracle DNN architecture.}


\subsubsection{Determining the reduced search space}

\hl{Using the detected values of the matrix parameters in Section~{\ref{sec:eval_nn}}, we first determine the possible connections between layers by locating shortcuts/branches and non-sequential connections.
For each possible connection configuration, we calculate the possible hyper-parameters for each layer.
The final search space is computed as}
\begin{equation}
search\ space = \sum_{i=1}^{C}{(\prod_{j=1}^{L}{x_j}})
\end{equation}
\hl{where $C$ is the total number of possible connection configurations, $L$ is the total number of layers, and $x_j$ is the number of possible combinations of hyper-parameters for layer $j$.}

\begin{figure}[ht]
    \centering
    \includegraphics[width=.65\columnwidth]{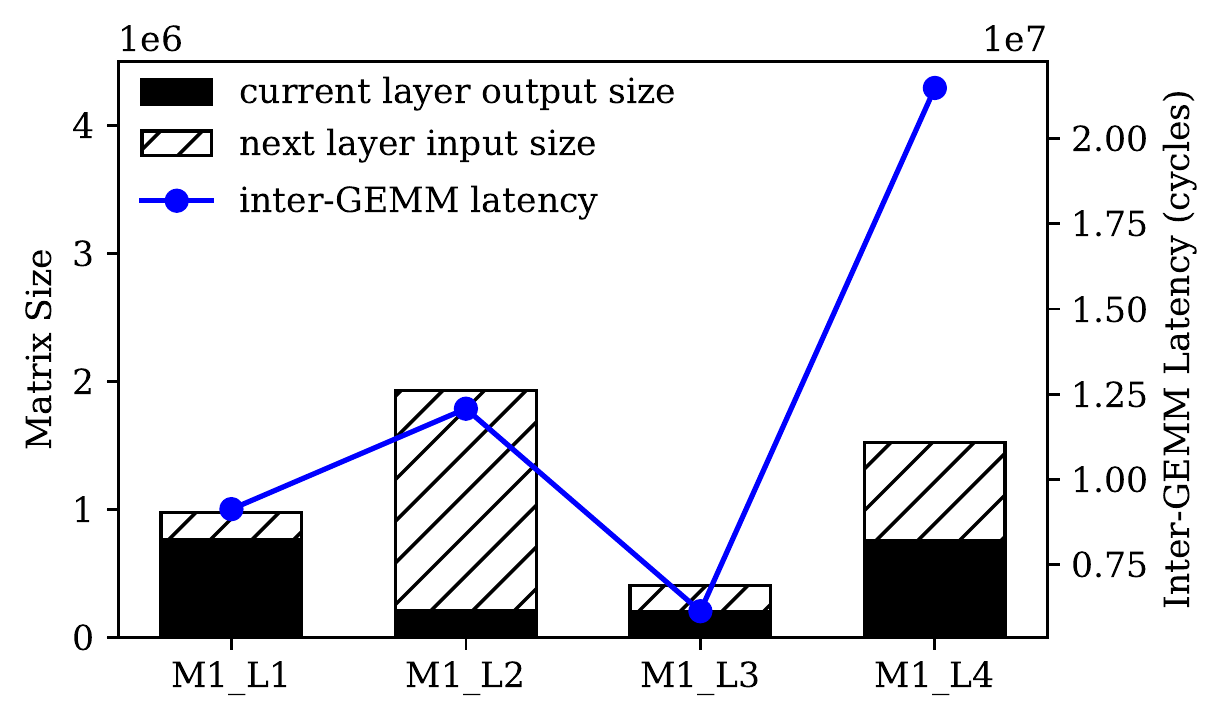}\hspace{15pt}\includegraphics[width=.2\columnwidth]{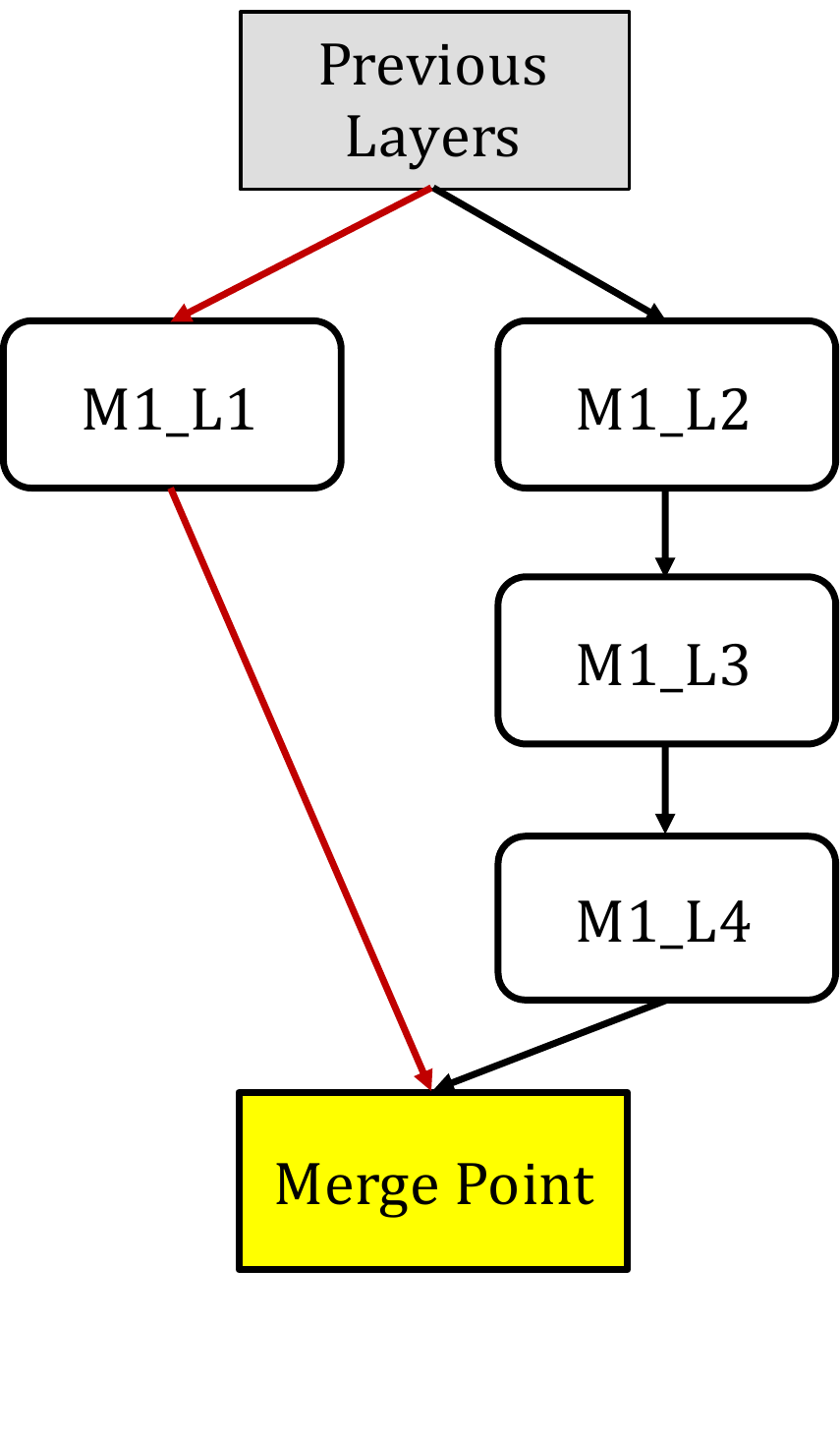}\vspace{-10pt}
    \caption{\hl{Extracting connections between layers in ResNet\_M1.}}
    \label{fig:inter_gemm}
\end{figure}

\paragraph{\hl{Determining connections between layers}}
\hl{We first try to determine the connections between layers.
As discussed in Section~{\ref{sec:map_shortcut}}, we can leverage the inter-GEMM latency to determine the sink of a shortcut or a branch.
Figure~{\ref{fig:inter_gemm}} shows the extracted input and output matrix sizes, and the inter-GEMM latencies for the layers in ResNet\_M1.
The inter-GEMM latency for M1\_L4 is significantly larger than expected, given its corresponding input and output matrix sizes, and thus can be identified as a sink.
Using the extracted dimension information (values of $n$ and $m$) in Figure~{\ref{fig:dnn_gemm}}, we determine that only M1\_L1 matches the output matrix size of M1\_L4. Therefore, we know that 
M1\_L1 is the source.
Further, by applying the DNN constraints on consecutive layers (Section~{\ref{sec:map_shortcut}}), we find that M1\_L1 and M1\_L2 are not consecutively connected, because $m_{M1\_L2}$ cannot be the product of $n_{M1\_L1}$ and the square of an integer.}


\hl{Based on the analysis above, we reverse engineer the
connections among the 4 layers, as
shown in Figure~{\ref{fig:inter_gemm}}. These connections
exactly match the actual ones in ResNet.
We use the same method to derive the possible 
connection configurations for the other modules.}

\paragraph{\hl{Determining hyper-parameters for each layer}}
\hl{We plug the detected matrix parameters into the formulas in Table~{\ref{tab:parameters}} to deduce the hyper-parameters for each layer.
For those matrix parameters that cannot be extracted precisely, we  leverage DNN constraints to prune the search space, as discussed in Section~{\ref{sec:map_shortcut}}.
For example, for convolutional layers, we use the $n$ and $k$ dimensions between two consecutive layers to deduce the number of filters and the filter sizes.}

\hl{As an example, consider L3 in ResNet\_M3.
First, we round the extracted $n$ to the nearest multiple of 64 to get 512, which is the number of filters in that layer. Since we only
have a very small error when detecting the $n$ dimension, we always obtain the correct number of filters in Figure~{\ref{fig:dnn_gemm}}(a).
Second, we use the formula in Table~{\ref{tab:parameters}} to determine the filter size.
The value for $n_{M2\_L2}$ is $118$,
but we cannot determine the value for $k_{M2\_L3}$.
As discussed in Section~{\ref{sec:map_shortcut}},
for consecutive layers, the $k$ dimension of the current layer must 
be the product of the $n$ dimension of the last layer and 
the square of an integer.
Considering that $k_{M2\_L3}$ is within the range $[68,384]$, the only possible value is $118$. In this way, we successfully reduce the parameter range to a single value, and deduce that the filter size is 1.}

\hl{We apply the same methodology for the other layers.
With this method, we obtain the solid circles in 
Figure~{\ref{fig:dnn_gemm}}. In some cases, this methodology 
generates two possible deduced values. For example,
this happens for the $k$ parameter in ResNet\_M3 Layer 4.
In this case, we have two solid circles in 
Figure~{\ref{fig:dnn_gemm}}, and we have to consider
both architectures.}

\paragraph{\hl{Determining pooling/striding}}
\hl{We use the differences in the $m$ dimension between 
consecutive layers to determine the pool or stride size.
For example, we see that the $m$ dimension changes from ResNet\_M1 to ResNet\_M2.
Even though we cannot precisely determine the value of $m_{M2\_L1}$, we can use DNN constraints to reduce the possible number of pooling/striding sizes.
As discussed in Section~{\ref{sec:map_shortcut}}, the reduction in the $m$ dimension must be the square of an integer.
Since $m_{M1\_L4}=3072$ and $m_{M2\_L1}$ is within the range $[524,1536]$, we can deduce that $m_{M2\_L1}$ is $768$ and that
the pool/stride size is $2$.
Since we cannot distinguish between pool layers and stride operations, 
we still have two possible configurations in the final search space:
one with a pool layer, and one with a stride operation.}

\subsubsection{The size of the reduced search space}

\hl{Based on the previous discussion,
Table~{\ref{tab:space_size}} shows the original size 
of the search space, and the reduced one after using \arch.
Recall that we calculate the original space assuming 
that the attacker already knows the total number of layers
and the type of layers, 
which is a conservative assumption.}


\begin{table}[t]
    \centering
    \footnotesize
    \begin{tabular}{|c||c||c|c|}
    \hline
      DNN    & Original: No  & \multicolumn{2}{c|}{Using \arch} \\\cline{3-4}
          & Side Channel &OpenBLAS & MKL \\\hline\hline
     ResNet-50  & $>5.8\times10^{46}$  & 512 & 6144\\\hline
     VGG-16  & $>4\times10^{35}$ &16 & 64 \\
    \hline
    \end{tabular}
    \vspace{-5pt}
    \caption{\hl{Comparing the original and reduced search spaces.}}
    \label{tab:space_size}
\end{table}

\hl{Using \arch, we are able to significantly reduce the search space from an intractable size to a reasonable size.
The reduced search space
is smaller in 
the VGG network because of its relatively small number of layers. Further, our attacks on MKL are less effective than on OpenBLAS.
This is because the matrix sizes in ResNet\_M4 are small, and MKL handles these matrices specially (Section~{\ref{sec:mkl}}).
Even after applying DNN constraints, we can only limit the number of possible values for some parameters in each layer of
that module to $2$ or $4$, which leads to the wider space size.}

\section{Countermeasures}
\label{sec:countermeasures}


We overview possible countermeasures against our attack, and 
discuss their effectiveness and performance implications.

Since our attack targets BLAS libraries, we first investigate whether it is possible to stop the attack by modifying the libraries.
One approach is to use less aggressive optimization.
It is unfeasible to abandon blocking completely, since unblocked matrix multiplication has poor cache performance.
However, it is possible to use a less aggressive blocking, with tolerable performance degradation. 
Specifically, we can remove the optimization for the first
iteration of Loop 3 (Lines 4-7 in Algorithm~\ref{algo:openblas_gemm}).
In this case, it will be difficult for the attacker to precisely recover the values of $n$ and $k$ without a much more detailed timing analysis. 

\hl{Another approach is to reduce the sizes of the 
dimensions of the matrices.
If the sizes of two or more dimensions of a matrix are smaller than the
block size, the attacker can only obtain ranges for the 
values of those dimensions. This makes the attack much harder.
There are existing techniques to reduce the sizes of the 
dimensions of a matrix.
For example, linear quantization reduces the 
weight and input precision, which 
means that the matrix becomes smaller. 
This mitigation is typically effective for the last few layers in a convolutional network, which use  relatively small matrices. 
However, it cannot protect layers with large matrices,
such as those using a large number of filters and input activations.}




Alternatively, one can use 
existing cache-based side channel defense mechanisms. One approach
is to disallow resource sharing. \hl{For example, the
MLaaS provider can disallow server sharing between different users. 
However, this is an extreme solution with major 
disadvantages in throughput. Another approach is to disable 
page sharing and page de-duplication. Unfortunately, while this defeats
the Flush+Reload attack, it does not defeat Prime+Probe.}




Another approach is to use cache partitioning, such as 
Intel CAT (Cache Allocation Technology), which assigns 
different ways of the LLC to different applications~\cite{catalyst}.
Since attackers may also target data access patterns, 
both code and data need to be protected in the cache. This means that the 
GEMM block
size needs to be adjusted to the reduced number of available 
ways in the LLC.
There will be some performance degradation due to reduced 
LLC capacity, but this is likely a good trade-off between 
performance and security.


\hl{Further, there are proposals for security-oriented cache
mechanisms such as PLCache~{\cite{plcache}}, Random Fill Cache~{\cite{randomfill}}, SHARP~{\cite{sharp}} and SecDCP~{\cite{secdcp}}.
If these mechanisms are adopted in production hardware, they 
can mitigate our attack with moderate performance degradation. Finally, one can add security features to performance-oriented cache partitioning proposals~{\cite{pipp}},{\cite{tadip}},{\cite{ucp}},{\cite{aggressor}}, to achieve better trade-offs between security and performance.}




\section{Related Work}\label{sec:related}


Recent research has called attention to the confidentiality of ML hyper-parameters.
Hua et al.~\cite{reverse_cnn_dac} designed the first attack to steal CNN architectures running on a hardware accelerator. Their attack is based on a different threat model, which requires the attacker to be able to monitor all of the memory accesses issued by the victim, including the type and address of each access. Our attack does not require such elevated privilege.

Cache-based side channel attacks have been used to trace program execution to steal sensitive information.
However, there is no existing cache-based side channel attack that can achieve the goal of extracting all the hyper-parameters of a DNN,
given the complexity of DNN computations, and the multi-level loop structure of GEMM,
as well as the high number of parameters to extract.
Most side channel attacks target cryptography algorithms, such as 
AES~\cite{gullasch2011cachegame,accessAES, sa},
RSA~\cite{liu2015practical,yarom2014flushreload} and ECDSA~\cite{garcia2016constant}.
These attacks directly correlate binary bits within the secret key with either branch execution
or matrix location access. 
Our attack on GEMM requires careful selection of probing addresses, and complicated post-analysis.
Furthermore, our attack can extract a much higher number of hyper-parameters, instead of a secret key.

\hl{GEMM is an important application in high-performance computing,
and there is a popular line of research on improving the
memory hierarchy performance of GEMM (e.g.,~{\cite{software_datacache}},{\cite{memory_systems}},{\cite{impulse}}).}

\section{Conclusion}\label{sec:conclusion}

In this paper, we proposed \arch: a fast and accurate mechanism to steal a DNN's architecture using cache-based side channel attacks.
We identified that DNN inference relies heavily on blocked
GEMM, and provided a detailed security analysis of this operation.
We then designed an attack to extract the input matrix 
parameters of GEMM calls, and scaled this attack to complete DNNs.
We used Prime+Probe and Flush+Reload to
attack VGG and ResNet DNNs running OpenBLAS and Intel MKL libraries. 
Our attack is effective in helping obtain the architectures
by very substantially reducing the search space of target 
DNN architectures. For example, for VGG using OpenBLAS,
it reduced the search space from more than $10^{35}$ 
architectures to just 16.


\bibliographystyle{ieeetr}
\bibliography{reference}

\end{document}